\documentclass[aps,pra, twocolumn,nofootinbib, showpacs]{revtex4-1} 
\usepackage[unicode=true,pdfusetitle,
bookmarks=true,bookmarksnumbered=false,bookmarksopen=false,
breaklinks=false,pdfborder={0 0 0},backref=false,colorlinks=false] {hyperref}
\hypersetup{ colorlinks,linkcolor=myurlcolor,citecolor=myurlcolor,urlcolor=myurlcolor}
\usepackage{braket,colortbl,amsthm,amsmath,cleveref,amssymb,txfonts}\definecolor{myurlcolor}{rgb}{0,0,0.7}
\usepackage{graphics,graphicx}
\usepackage{color}
\usepackage{amssymb}
\usepackage{amsthm}
\usepackage{amsfonts}
\usepackage{float}
\usepackage{graphicx}
\usepackage[format = plain,labelfont = bf,up, textfont = normal , up, justification
=raggedright, singlelinecheck =false]{caption}
\usepackage{subcaption}
\usepackage{tabularx}
\usepackage{amsmath}
\usepackage{braket}

\usepackage{graphicx}
\usepackage[utf8x]{inputenc}
\usepackage{color,soul}
\usepackage{amsmath}
\usepackage{braket}
\usepackage{latexsym}
\usepackage{amssymb}
\usepackage{amsthm}
\usepackage{xcolor}
\usepackage{bm}
\usepackage{graphics,epstopdf}
\usepackage{color}
\usepackage{enumitem}
\usepackage{fmtcount}
\usepackage{booktabs}
\usepackage{epsfig}
\usepackage[normalem]{ulem}
\theoremstyle{plain}
\newtheorem{theorem}{Theorem}
\newtheorem{cor}[theorem]{Corollary}
\def\bea{\begin{eqnarray}}
\def\eea{\end{eqnarray}}
\def\ba{\begin{array}}
\def\ea{\end{array}}
\def\n{\nonumber}
\def\ket{\rangle}
\def\bra{\langle}
\def\T{\text{T}}

\def\Tr{\text{Tr}}
\def\ts{\tau_s}
\def\ot{\omega_t}

\begin{document}
\title{Noisy quantum input loophole in measurement-device-independent entanglement witnesses}

\author{Kornikar Sen, Chirag Srivastava, Shiladitya Mal, Aditi Sen(De), Ujjwal Sen}

\affiliation{Harish-Chandra Research Institute, HBNI, Chhatnag Road, Jhunsi,
Allahabad 211 019, India}
\begin{abstract}
Entanglement witnesses form an effective method to locally detect entanglement in the laboratory without having the prior knowledge of the full density matrix. However, separable states can be erroneously indicated as entangled in such detections
in the presence of wrong measurements or loss in detectors. Measurement-device-independent entanglement witnesses (MDI-EWs) never detect fake entanglement even under wrong measurements and for a particular kind of lossy detectors. A crucial assumption in the case of faithful detection of entanglement employing MDI-EWs is that the preparation devices  producing ``quantum inputs'' -- which are inputs additional to the quantum state whose entanglement is to be detected -- are perfect and there is no noise during their transmission. Here, we relax these assumptions and provide a general framework for studying the effect of noise on the quantum inputs, invoking uniform and non-uniform noise models. We derive sufficient conditions on the uniform noisy map for retaining the characteristic of MDI-EWs. We find that in the context of non-uniform and entangling noise, fake entanglement detection is possible even by MDI-EWs. We also investigate various paradigmatic models of local noise and find conditions of revealing entanglement in the class of Werner states.
\end{abstract}
\maketitle

\section{Introduction}
\label{sec1}
 A most fascinating characteristic of a composite quantum state is that it may be in a state where full knowledge about the whole does not allow to have full knowledge about it parts, which - in  case when the whole state is pure - is called entanglement \cite{Hor'09,Guhne'09,Das'19}. Entanglement can also exist in mixed states, and over the last thirty years or so, several tasks, like teleportation 
\cite{Ben'93}, dense coding \cite{Ben'92},  secure key distribution \cite{Ekert'91}, and measurement-based  computation 
\cite{Lagnajita}
have been devised, which have been performed using quantum entanglement as resource with greater efficiency than their classical analogs. Unsurprisingly therefore, in this era of the so-called second-generation quantum technologies, detection of entanglement is gaining more and more significance day by day, both theoretically and experimentally.
For example, given the description of a shared state, one can apply the positive partial transpose criterion \cite{Peres'96,Hor'96}, range criterion \cite{Hor'97}, and some other criteria \cite{Hor'09, Guhne'09, Das'19} to find whether the state is entangled, whereas violation of Bell inequality \cite{Bell} and other entanglement witnesses (EWs) \cite{Hor'96,Guhne'09,terhal'00} are employed in the laboratory, respectively when there is no and partial knowledge of the shared state.

In entanglement theory, an important idea is that given any entangled state, there always exist a hermitian witness operator such that its expectation value for the given state is negative, while the same is non-negative for all separable states \cite{Hor'96,terhal'00}. Moreover, these witness operators are implementable via local measurements. An important drawback of this method is that a separable state may turn out to be entangled to an experimentalist if wrong measurements were performed \cite{See'01,Bruss'07,See'07,Moro'10,Ban'11,Ros'12}. Thus, if the measurement devices are not trusted, the standard EWs may not guarantee entanglement.
On the other hand, violation of Bell inequality certifies entangled states independent of measurements being performed \cite{Ban'11, Mayers'04, Mckague'12}. However, witnessing entanglement via violation of Bell inequality does not always provide an appropriate method of identifying entanglement as there are entangled states not exhibiting ``Bell-nonlocality'' \cite{werner,barrett}. 

To overcome the measurement dependence of standard EWs, the concept of measurement-device-independent entanglement witnesses (MDI-EWs) were introduced  \cite{Cyril'13}. Specifically, based on the idea of a semi-quantum nonlocal game \cite{Bus'12}, it was shown that one can  construct an MDI-EW from a known standard EW \cite{Cyril'13}. Unlike the ``Bell nonlocal game'' \cite{Cleve'04} for a bipartite scenario, each experimenter gets quantum inputs from a ``referee'', instead of classical inputs, in a semi-quantum nonlocal game \cite{Bus'12}, and in this extended nonlocal scenario, all entangled states turn out to be more ``resourceful'' than all separable states. The advantage of measurement-device independence was also explored in the context of quantum key distribution \cite{Lo'12}, verification of quantum steering phenomena \cite{steering}, multiple sharing of bipartite entanglement \cite{Chirag'19}, etc.

It is interesting to mention here that the witnesses of entanglement are not always free from  loopholes.
It was known previously that the Bell test suffers from the detection loophole, and methods to address the problem was much discussed \cite{Bruss'07,Det}. Though the MDI-EWs do not mistake a separable state as entangled under wrong measurements, and are robust against a particular kind of lossy detectors, it is not free from the  detection loophole in general 
\cite{korni'20}. In all the studies so far regarding MDI-EWs, it was assumed that the sources preparing the quantum inputs are perfect. It is very natural to trust the preparation device over the testing device. Still, imperfection in the preparation device cannot be avoided completely. In the present work, we address the question how MDI-EWs get affected when imperfection in devices preparing the relevant quantum inputs is allowed.

More specifically, in this article, we explore the effect of noisy inputs on detection of entanglement in the measurement-device-independent way, when no other loopholes are present. We adopt a general framework of dealing with noisy maps on the quantum inputs, which can be characterized in two broad categories - (1) \emph{uniform noise} and (2) \emph{non-uniform noise}. If all the quantum inputs, provided by the ``referee'' to each party, are transmitted by the same type of noisy channel, then the situation is deemed as within the category of uniform noise.  Otherwise, it falls under the  category of non-uniform noise.
In the case of uniform noise, we derive a sufficient condition on the noisy channels such that the MDI-EWs retain their character of detecting entanglement. Specifically, if the adjoint map of the noisy channel takes any separable operator to a separable one, the MDI-EW never mistakes a separable state as an entangled one. We also present various examples of non-uniform noise models, including entangling maps, and, interestingly, find that it is possible to have false detection of entanglement using MDI-EWs. We then investigate conditions on detection of entanglement in Werner states \cite{Werner}, under various uniform noise models, viz., white and colored noise, bit-flip, phase-flip, bit-phase flip, and the amplitude damping channel. Finally, the effect of a correlated noise model, in which the noises over the quantum inputs of both the parties are classically correlated \cite{vello'11}, is also addressed.

This rest of the paper is organized as follows. In Sec. \ref{sec2}, we briefly review the MDI-EWs and discuss their modified forms under noisy quantum inputs. In Sec. \ref{sec3}, a sufficient condition on the uniform noisy channel is derived such that no fake entanglement detection will occur. In Sec. \ref{sec4}, two examples of noisy maps are presented which may indicate a separable state as an entangled one. In Sec. \ref{sec5}, we obtain the conditions of detecting an entangled state under various noises, and we end with concluding remarks in Sec. \ref{sec6}.

\section{Measurement-device-independent entanglement witness}
\label{sec2}
In this section, we briefly review the construction of an MDI-EW following Ref.~\cite{Cyril'13}.   
An EW operator, $W$, which is a hermitian operator acting on the Hilbert space,  $\mathcal{H}_A\otimes\mathcal{H}_B$, of dimension $d_A \times d_B$, can be decomposed in the following way:
\begin{equation}
W=\sum_{s,t}\beta_{s,t}\tau_s^\T \otimes \omega_t^\T. \label{eq1}
\end{equation}
Here $\{\tau_s^\T \}$ and $\{\omega_t^\T \}$ are sets of density matrices acting on $\mathcal{H}_A$ and $\mathcal{H}_B$ respectively, and 
$\beta_{s,t}$ is a set of real numbers. The superscript ``T" over $\tau_s$ or $\omega_t$ denotes their transpositions. This decomposition is not unique. Suppose now that Alice and Bob share a bipartite quantum state, \(\rho_{AB}\), on \(\mathcal{H}_A\otimes\mathcal{H}_B\). Additionally, Alice and Bob are provided respectively with the sets of states $\{ \ts  \}$ and $\{\ot \}$ which satisfy Eq. \eqref{eq1}.
Alice performs a positive operator-valued measurement (POVM), on her part of the shared state, $\rho_{AB}$, and a randomly obtained state from $\{ \ts  \}$. Similarly, Bob also measures jointly on his part of $\rho_{AB}$ and a randomly obtained state from $\{\ot\}$. Let the POVM at Alice's (Bob's) end be whether the projection onto a fixed maximally entangled state occurs or not, viz., 
$\{|\Phi^+_{A(B)}\ket\bra\Phi^+_{A(B)}|,I_{d^2_{A(B)}}-|\Phi^+_{A(B)}\ket\bra\Phi^+_{A(B)}|\}$, where $I_{d^2_{A(B)}}$ is the identity operator on the $d^2_{A(B)}$-dimensional complex Hilbert space, and $|\Phi^+_{A(B)}\ket=\frac{1}{\sqrt{d_A(d_B)}}\sum_{i=1}^{d_A(d_B)}|i\ket  \otimes |i \ket$, with the tensor product being between Alice's (Bob's) part of \(\rho_{AB}\) and her (his) randomly obtained state.
We denote the clicking of $|\Phi^+_{A(B)}\ket\bra\Phi^+_{A(B)}|$ as outcome `1', and thus the joint probability of getting `1' by both Alice and Bob is given by
\begin{equation}\label{bla}
P(1,1|\ts,\ot)=\Tr[(|\Phi^+_{A}\ket\bra\Phi^+_{A}|\otimes |\Phi^+_{B}\ket\bra\Phi^+_{B}|)(\ts \otimes \rho_{AB} \otimes \ot)]. 
\end{equation}
With this notation, the MDI-EW is defined as
\begin{equation}
I(P)=\sum_{s,t}\beta_{s,t}P(1,1|\ts,\ot).
\end{equation} 
Now it is straightforward to see that the MDI-EW is related to the standard EW in Eq. \eqref{eq1} as
\begin{equation}
	I(P)=\frac{\Tr(W\rho_{AB})}{d_A d_B}. \label{eq4}
\end{equation}
Hence, if $\Tr(W\rho_{AB})<0$, the corresponding $I(P)$ is also negative and is non-negative for all separable states. 

For completeness, let us first show that performing wrong measurements does not lead to fake entanglement detection, as is already known in the literature.

\subsection*{Measurement-device-independence of MDI-EW}
If instead of the measurement,  $\{|\Phi^+_{A(B)}\ket\bra\Phi^+_{A(B)}|,I_{d^2_{A(B)}}-|\Phi^+_{A(B)}\ket\bra\Phi^+_{A(B)}|\}$, Alice (Bob) performs some arbitrary measurement $\{A_1(B_1),A_2(B_2)\}$.
Let $A_1(B_1)$ be the POVM element corresponding to the outcome `1' in the joint probability expression given in Eq. \eqref{bla}.
It is seen that still the MDI-EW function is positive for any separable state. To prove this, let us consider a product state, $\sigma=\sigma_A \otimes \sigma_B$  acting on the Hilbert space $\mathcal{H}_A\otimes \mathcal{H}_B$. Then the corresponding MDI-EW can be written as
\begin{eqnarray}\label{pawn}
I(P_\sigma)=  \sum_{s,t}\beta_{s,t}\Tr[(A_1\otimes B_1)(\ts \otimes \sigma_A \otimes \sigma_B \otimes \ot)] \n \\
=\sum_{s,t}\beta_{s,t}\Tr[(A_1'\otimes B_1')(\ts \otimes \ot)], 
\end{eqnarray}
where $A_1'=\Tr_s[A_1(I_A \otimes \sigma_A)]$ and $B_1'=\Tr_t[B_1(\sigma_B \otimes I_B)]$ are elements of effective POVMs acting on the quantum inputs $\ts$ and $\ot$ corresponding to outcomes `1'. Here, $I_{A(B)}$ denotes the identity operator on Hilbert space $\mathcal{H}_{A(B)}$ and $\Tr_i$ represents tracing over the pertaining input Hilbert space. Hence we get $I(P_\sigma)=\Tr[(A_1'^\T\otimes B_1'^\T)W]$. Since $A_1'^\T$ and $B_1'^\T$ are hermitian matrices, $I(P_\sigma )\geq 0$ for any POVM elements $A_1$ and $B_1$. Since $I(P_\sigma)$ is positive for any product state,  it is straightforward to see that it is positive for any separable state, whatever be the measurement performed.

\subsection*{MDI-EW for Werner states}\label{subsec2A}
To explain the construction of MDI-EW through a particular example, let us consider the Werner family \cite{Werner}, 
\begin{equation}
\rho_v=v|\Psi^-\ket\bra \Psi^- |+(1-v)I_4/4,
\end{equation}
 where $v\in [0,1]$, $|\Psi^-\ket=\frac{|01\ket-|10\ket}{\sqrt{2}}$, and $I_d$ is the $d$ dimensional identity operator. Using the positive partial transposition criterion it can be shown that $\rho_v$ is entangled for $v>\frac{1}{3}$ \cite{neg}, and this entanglement can be detected by the witness operator, $W=\frac{1}{2}I_4-|\Psi^-\ket \bra \Psi^-|$ \cite{terhal'00,Guhne'09}. This witness operator can be decomposed as in Eq. \eqref{eq1} with
\begin{eqnarray}
	\beta_{s,t}=\frac{5}{8}\text{ if }s=t, \ \ \ \ \ \ \ \beta_{s,t}=-\frac{1}{8} \text{ if }s\neq t, \n \\
	\ts=\sigma_s\frac{I_2+\vec{n}.\vec{\sigma}}{2}\sigma_s, \ \ \ \ \ \ \
	\ot=\sigma_t\frac{I_2+\vec{n}.\vec{\sigma}}{2}\sigma_t. 
\end{eqnarray}  
Here, $s,t=0,1,2,3$, $\vec{n}=(1,1,1)/\sqrt{3}$, $\vec{\sigma}=(\sigma_1,\sigma_2,\sigma_3)$, i.e., the three Pauli matrices, and $\sigma_0=I_2$. Using this witness operator, $W$, one can construct the corresponding MDI-EW as
\begin{equation}
\nonumber
	I(P)=\frac{5}{8}\sum_{s=t}P(1,1|\ts,\ot)-\frac{1}{8}\sum_{s\neq t}P(1,1|\ts,\ot),
\end{equation}	
which, for the Werner state, leads to 
\begin{equation}
\nonumber I(P)=\frac{1-3v}{16}. 
\end{equation}

\subsection*{Modification of MDI-EW with noisy quantum inputs}

So far, we have assumed that the sources of input states are perfect and that there is no noise in the transmission lines which carry quantum inputs to the local experimenters. Let us now consider a situation where quantum inputs are not the desired ones. Let us describe the process within a general framework i.e., let us assume that the inputs are affected by some completely positive trace preserving (CPTP) map to create $\Lambda^{st}(\ts \otimes \ot)$, where $\Lambda^{st}$ is the CPTP map acting on the original inputs. The corresponding noisy MDI-EW is then given by
\begin{equation}
	I_{\Lambda^{st}}(P)=\sum_{s,t}\beta_{s,t}P(1,1|\Lambda^{st}(\ts \otimes \ot)). 
\end{equation}
Note that the noisy map, $\Lambda^{st}$, may depend on the quantum inputs, as indicated by superscript $st$. But if the noise on all input states are the same, i.e., $\Lambda^{st}$ has no explicit dependence on the indices $s$ and $t$, then we call such noise as \textit{uniform noise} over the quantum inputs, and otherwise, we call it \textit{non-uniform}. 

The modified MDI-EW under uniform noise, $\Lambda(.)$, is given by 
\begin{eqnarray}
	I_\Lambda(P)&=&\sum_{s,t}\beta_{s,t}\Tr[(|\Phi^+_{A}\ket\bra\Phi^+_{A}|\otimes|\Phi^+_{B}\ket\bra\Phi^+_{B}|)\n \\ \ && \ \ \ \ \ \ \ \ \ \ \ \ \times( \Lambda(\ts\otimes\ot)\otimes \rho_{AB})]\n \\
	&=&\sum_{s,t}\beta_{s,t}\Tr[ \Lambda(\ts\otimes\ot)^\T\rho_{AB}]/(d_Ad_B) \n \\
	&=&\Tr[W'\rho_{AB}]/(d_Ad_B), \label{naam}
\end{eqnarray}
where $W'=\sum_{s,t}\beta_{s,t}\Lambda(\ts\otimes\ot)^\T.$

In the next section, we discuss the effect of noisy inputs on the entanglement detection process.

\section{Status of separable states under MDI-EW with uniform noise on quantum
inputs}
\label{sec3}

Let us first consider the case where inputs are affected by uniform noise, and investigate whether the modified MDI-EW shows all the separable states as separable for arbitrary measurements. 
In the following theorem, we answer to this 
question for uniform noise on all quantum inputs. 
 
Before stating the theorem, we mention the definition of adjoint of a map. Let $O_1$ and $O_2$ be positive semidefinite operators on a Hilbert space and $M$ be a CPTP map  on the space of operators. Then the adjoint map, $M^+$, is defined by $\text{Tr}[O_1.M(O_2)]=\text{Tr}[M^+(O_1).O_2].$

\begin{theorem}
A uniform noisy map $\Lambda(.)$ is acting on the joint quantum inputs, $\tau_s \otimes \omega_t$, for all values of $s$ and $t$ in an MDI-EW setup. $I_\Lambda(P) \geqslant 0$ holds for all separable states and arbitrary measurements, if the adjoint map $\Lambda^+$ takes any separable positive semidefinite operators to another such operator. 
\end{theorem}
\begin{proof}
Let Alice and Bob share a product state, $\sigma=\sigma_A \otimes \sigma_B$. Thus the modified MDI-EW with effective POVM elements $A'_1=\Tr_s[A_1(I_A \otimes \sigma_A)]$ and $B'_1=\Tr_t[B_1(\sigma_B \otimes I_B)]$ (mentioned in Sec. \ref{sec2}) on the noisy inputs $\Lambda(\ts \otimes \ot)$ corresponding to both obtaining outcomes `1' is given by
\begin{eqnarray}\label{Ilambda}
I_{\Lambda}(P_\sigma)&=\sum_{s,t}\beta_{s,t}\text{Tr}[A'_1 \otimes B'_1\Lambda(\ts \otimes \ot)]. \n \\
&=\sum_{s,t}\beta_{s,t}\text{Tr}[\Lambda^+ (A'_1 \otimes B'_1)\ts \otimes \ot].
\end{eqnarray}
Since $\Lambda^+(.)$ has the property that it maps any product positive semidefinite operator to a separable operator on the Hilbert space $\mathcal{H}_A\otimes\mathcal{H}_B$, we can write
\begin{equation}
\Lambda^+ (A'_1 \otimes B'_1)=\sum_k \mathcal{A}_k\otimes \mathcal{B}_k,
\end{equation}
where $\mathcal{A}_k$ and $\mathcal{B}_k$ are positive semidefinite operators on $\mathcal{H}_A$ and $\mathcal{H}_B$ respectively. This implies that
\begin{equation}
 I_{\Lambda}(P_\sigma)=\sum_k\sum_{s,t}\beta_{s,t}\text{Tr}[(\mathcal{A}_k\otimes \mathcal{B}_k)(\ts \otimes \ot)]\geqslant0,
\end{equation}
as argued in Sec. \ref{sec2}, in the proof of $I(P_\sigma)\geqslant 0$. Since $I_\Lambda(P_\sigma)\geqslant 0$  for any product state, it remains non-negative for arbitrary convex combinations of any product states, i.e.,  for all separable states.  
\end{proof}
Note that if the noisy CPTP map, $\Lambda(.)$, is dependent on indices $s$ and $t$,  the argument used in Sec. \ref{sec2} will not be applicable. Actually, we give an example in the next section, in which a map dependent on the indices, or in other words, a non-uniform noisy map, leads to a separable state  being mistaken as entangled by the noisy MDI-EW. Note that a noisy map of the form $\Lambda(\ts\otimes\ot)=\Lambda_1(\ts)\otimes\Lambda_2(\ot)$ is called local noise.
Moreover,  uniform local noisy maps form a subset of uniform noisy maps taking separable operators to separable operators. 
We, therefore, have the following corollary.
\begin{cor} The modified MDI-EW under a local uniform noise on quantum inputs  never indicates a separable state as an entangled one, provided the adjoints of the local noises preserves the positive semidefiniteness of operators.
\end{cor}

\section{Fake entanglement detection under non-uniform and Global noise}
\label{sec4}
In this section, we discuss two examples of noise models on quantum inputs, for which the noisy MDI-EWs can mistake  separable states 
as
entangled. In both the examples, we stick to the MDI-EWs corresponding to the Werner class of states, $\rho_v$, mentioned in Sec. \ref{subsec2A}.\\ 

\noindent \textbf{Example 1.} \textit{A class of non-uniform local noisy channels on the quantum inputs}:\\
Let the quantum inputs, after getting affected by such noise process, transform into
\begin{eqnarray}
\label{nonuni}
\ts'=p_s^A\ts+(1-p_s^A)|\theta\ket\bra\theta|,\n\\
\ot'=p_t^B\ot+(1-p_t^B)|\theta\ket\bra\theta|,
\end{eqnarray}
where $\ts$ and $\ot$ for $\rho_v$ are given in Sec.~\ref{subsec2A},  $|\theta\ket$ is a pure state in the Hilbert space of dimension two, and $p_s^A$ and $p_t^B$ are probabilities. Notice that Eq.~\eqref{nonuni} is an example of non-uniform noise over the quantum inputs, and the joint noisy quantum input can be written as $\ts' \otimes \ot' = \Lambda^{st}(\ts \otimes \ot)$.

We will now show that there exist measurements such that the modified MDI-EW becomes negative, i.e.,  $I_{\Lambda^{st}}(P_\sigma)<0$, due to the given noisy quantum inputs $\ts'$ and $\ot'$, even for a product state $\sigma=\sigma_A \otimes \sigma_B$. 
Let $A'_1=\Tr_s[A_1(I_A \otimes \sigma_A)]$ and $B'_1=\Tr_t[B_1(\sigma_B \otimes I_B)]$ be elements of the effective POVM acting on $\Lambda^{st}(\ts \otimes \ot)$ corresponding to outcomes `1' on both sides. Since the effective POVM elements can be arbitrary, we choose,  without loss of generality, $A'_1=B'_1=|\theta^{\perp}\ket\bra\theta^{\perp}|$, where $|\theta^{\perp}\ket$ is orthogonal to $|\theta\ket$.
Therefore, the noisy MDI-EW is
\begin{equation}
I_{\Lambda^{st}}(P_\sigma)=\sum_{s,t}\beta_{s,t}p_s^Ap_t^B \bra\theta^{\perp}|\ts|\theta^{\perp}\ket  \bra\theta^{\perp}|\ot|\theta^{\perp}\ket. 
\end{equation}

Consider now the following choice of probabilities:
\begin{center}
 \begin{tabular}{| c | c | c |} 
 \hline
 $s$/$t$ & $p_s^A$ & $p_t^B$  \\ [1ex] 
 \hline\hline
 0 & q & 0  \\ 
 \hline
 1 & q & 0 \\
 \hline
 2 & q & 0 \\
 \hline
 3 & 0 & q \\[1ex]
 \hline
\end{tabular}
\end{center}
Here, $0\leqslant q\leqslant 1$. Now with this choice of probabilities, we obtain
\begin{equation}
I_{\Lambda^{st}}(P_\sigma)=\frac{-q^2}{8}\bra\theta^{\perp}|\omega_3|\theta^{\perp}\ket\sum_{s=0}^2 \bra\theta^{\perp}|\ts|\theta^{\perp}\ket. 
\end{equation}
Note that the expectation value of any quantum state is non-negative.
Thus the quantity, $I_{\Lambda^{st}}(P_\sigma)$, can either be zero or negative, but, since the pure state $|\theta\ket$ can be arbitrarily chosen, therefore, there always exist several choices of $|\theta\ket$ (actually, an  infinity of them) such that $I_{\Lambda^{st}}(P_\sigma)<0$.
Notice that $\beta_{s,t}=\frac{-1}{8}$, if $s\neq t$, and equal to $\frac{5}{8}$ when $s=t$. Therefore, the probabilities mentioned in the table are intentionally used such that no contribution for $I_{\Lambda^{st}}(P_\sigma)$ comes up from the case $s=t$.\\

\noindent \textbf{Example 2.} \textit{Noise mapping separable quantum inputs to entangled ones:}\\
Suppose now that the noisy quantum states coming to the observers $A$ and $B$, although initially separable at preparation, somehow becomes entangled on transmission, having reduced states $\ts'$ and $\ot'$, respectively. Such entanglement could also occur due to malicious referee who intentionally replaces a product input by an entangled one, to derail the entanglement detection process. The same could also happen due to a preparation procedure where the source preparing the inputs for Alice is not completely separated from the one that prepares the inputs for Bob. 
For example, consider $\ts'=p\ts+\frac{1-p}{2}I_2$ and $\ot'=p\ot+\frac{1-p}{2}I_2$. Note that the quantum inputs, $\ts$ and $\ot$, are pure states and thus will be denoted by $|\psi_s\ket$ and $|\psi_t\ket$ for this example. Now, 
let a global map over any pair of quantum inputs (each pair consists of one quantum input from Alice and another from Bob) be defined as  $\Lambda(\ts \otimes \ot)\equiv |\chi_{st}\ket\bra\chi_{st}|$, with $$|\chi_{st}\ket=\sqrt{\frac{1+p}{2}}|\psi_s\psi_t\ket+\sqrt{\frac{1-p}{2}}|\psi^\perp_s\psi^\perp_t\ket, $$ where $|\psi^\perp\ket$ denotes the orthogonal qubit state (with an arbitrarily chosen phase) to an arbitrarily chosen qubit state, $|\psi\ket$. It is easy to see that
\begin{eqnarray}
\ts'=\text{Tr}_{t}[\Lambda(\ts \otimes \ot)], \n \\
\ot'=\text{Tr}_{s}[\Lambda(\ts \otimes \ot)].
\end{eqnarray}
Let us consider the scenario of verification of measurement-device independence for any MDI-EW. Let Alice and Bob share a product quantum state $\sigma_A \otimes \sigma_B$ and perform arbitrary measurements as considered to obtain Eq. \eqref{pawn}.
Therefore, the noisy MDI-EW using the quantum inputs $|\chi_{st}\ket$ for the given scenario takes the form,
\begin{equation}
I_\Lambda(P_\sigma)=\sum_{st}\beta_{st}\text{Tr}[A'_1\otimes B'_1.|\chi_{st}\ket\bra\chi_{st}|].\nonumber
\end{equation}
It can be checked that for $p=0$, $A'_1 = (I_2+\sigma_y)/2$ and $B'_1=(I_2-\sigma_y)/2$, $I_\Lambda(P_\sigma)$  takes a negative value, -0.041, correct to the third decimal place. Therefore, there exist separable states which can be detected as entangled under the situation of entangled quantum inputs, and hence measurement-device independence of the given noisy MDI-EW is not valid.


\section{Detection of entanglement under various uniform noise models}
\label{sec5}

In Sec.~\ref{sec3}, we have shown that under special types of uniform local or global noise, 
no separable state will appear as entangled. In this section, we consider different types of uniform noise on the input states, none of which  leads to fake detection of entanglement for Werner states. We derive conditions for witnessing entanglement of Werner states using MDI-EWs.

\subsection{White noise: Admixture with completely depolarized state}

With admixture with white noise with a pure state, say $\xi$, the state becomes
\begin{equation}
\xi_{wn}(p)=p\xi+\frac{(1-p)}{2}I_2, \n
\end{equation}
where $p$ is the noise parameter, so that \(1-p\) is the amount of white noise that gets mixed with the original state. The subscript, $wn$, stands for ``white noise''. Let the input states that Alice and Bob get from the referee get admixed with white noise with noise parameters $p_1$ and $p_2$ respectively. The new (i.e., altered) input states read 
\begin{equation}
(\ts)_{wn}=p_1\ts+\frac{1-p_1}{2}I_2, \ \ \ \ \ \ \
(\ot)_{wn}=p_2\ot+\frac{1-p_2}{2}I_2. \n
\end{equation}
The corresponding noisy MDI-EW is given by
\begin{equation}
	I_{wn}(P)=\sum_{s,t}\beta_{s,t} P(1,1|(\ts )_{wn},(\ot )_{wn}). \n
\end{equation} 
Considering white noise in Eq. \eqref{naam}, we have
\begin{widetext}
\begin{equation}
	I_{wn}(P)=\frac{1}{d_Ad_B}\sum_{s,t}\beta_{s,t}\Tr\left[ p_1p_2\ts^\T\otimes\ot^\T+\frac{p_1(1-p_2)}{2}\ts^\T\otimes I_2+\frac{(1-p_1)p_2}{2}I_2\otimes \ot^\T+\frac{(1-p_1)(1-p_2)}{4}I_2\otimes I_2\right]. \n
\end{equation}
\end{widetext}

For the MDI-EW corresponding to the Werner states, $\sum_{s,t}\beta_{s,t}\ts=\sum_{s,t}\beta_{s,t}\ot=I_2/2$.
Hence, due to the presence of white noise, the noisy MDI-EW of an Werner state becomes 
\begin{eqnarray}
	I_{wn}(P)&=&\frac{1}{4}\Tr\left[\left(p_1p_2\sum_{s,t}\beta_{s,t}\ts^\T\otimes\ot^\T +\frac{1-p_1p_2}{4}I_4\right)\rho_{AB}\right] \n \\
	&=&p_1p_2I(P)+\frac{1-p_1p_2}{16}. \n
\end{eqnarray}
The above MDI-EW will be negative if
\begin{eqnarray}
	&&p_1p_2I(P)+\frac{1-p_1p_2}{16}<0\n \\
\Rightarrow	&&p_1p_2\left(\frac{1-3v}{16}\right)+\frac{1-p_1p_2}{16}<0 \n \\
\Rightarrow	&&v>\frac{1}{3p_1p_2}. \label{eq7}
\end{eqnarray}
So, if for the shared quantum state, $\rho_v$, the inequality in \eqref{eq7} is satisfied, $\rho_v$ is entangled and can be detected by the noisy MDI-EW, $I_{wn}(P)$.

\subsection{General admixture: Admixing with arbitrary quantum state}
 \begin{figure*}[t]
 	\includegraphics[scale=1.2]{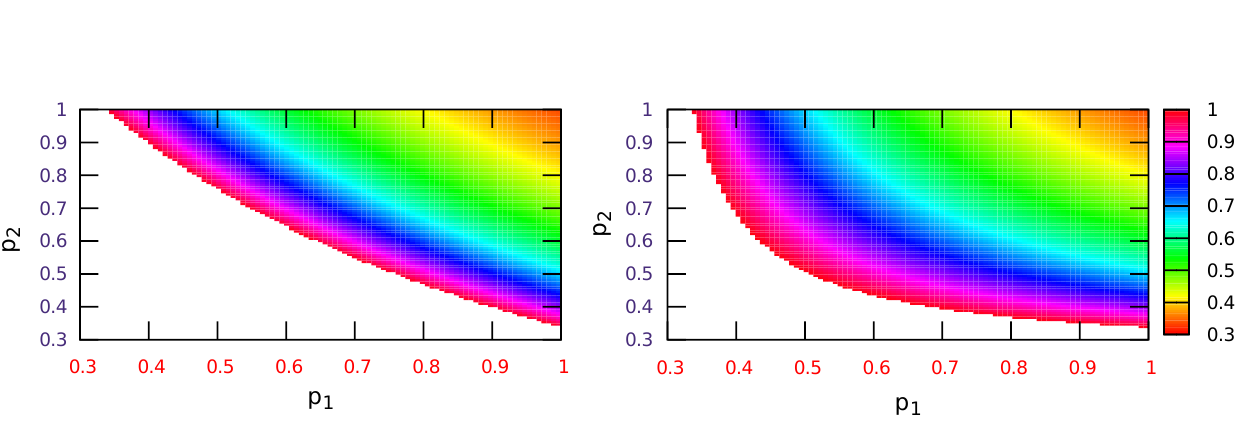}
 	\caption{Depiction of detecting entangled Werner states
using MDI-EW with Alice (Bob) using noisy quantum inputs in which ideal quantum inputs are mixed with quantum states \(X\) (\(Y\)), given in Eq. \eqref{rafi}. The Werner state parameter $\nu$ is lower bounded by a quantity $A^{-1}$ which is a function of states, $X$ and \(Y\), and the noise parameters, $p_1$ and $p_2$. The left panel correspond to the case when the quantum states $X$ and $Y$ are such that $A=A_{min}$, whereas, the right panel correspond to the case when $A=A_{max}$. The axes denote the corresponding noise parameters. An entangled state
would be detected if the values of the noise parameters are such that the
point belongs to the colored region and the value of $\nu$ is greater than the
number corresponding to that color indicated in the colored box. 
 All quantities are dimensionless.} 
 	\label{fig1}
 \end{figure*}
 
Consider now a more general noisy channel, viz. admixture of  arbitrary quantum states with the input states. The altered inputs are now 
\begin{equation}\label{rafi}
	(\ts)_{qs}=p_1\ts+(1-p_1)X\text{, }(\ot)_{qs}=p_2\ot+(1-p_2)Y. 
\end{equation} 
Here $X=\left[ \begin{matrix}
x_0&&x_1+ix_2\\
x_1-ix_2&&1-x_0
\end{matrix}\right]$ and $Y=\left[ \begin{matrix}
y_0&&y_1+iy_2\\
y_1-iy_2&&1-y_0
\end{matrix}\right]$ are density matrices and the subscript $qs$ in a noisy quantum input is used to indicate that it is a mixture of an ideal quantum input with any quantum state.  Note that since $X$ and $Y$ are quantum states, therefore, $x_0$, $x_1$, and $x_2$ satisfy $(1-2x_0)^2+(2x_1)^2+(2x_2)^2\leq 1$ and similarly  $y_0$, $y_1$, and $y_2$ satisfy $(1-2y_0)^2+(2y_1)^2+(2y_2)^2\leq 1$. Considering these noisy input states in the MDI-EW operator, 
we get
\begin{widetext}
\begin{eqnarray}
	I_{qs}(P)&=&\frac{p_1p_2}{4}\Tr[W\rho_v]+\frac{p_1(1-p_2)}{4}\Tr[I\otimes X^\T\rho_v]+\frac{p_2(1-p_1)}{4}\Tr[Y^\T\otimes I\rho_v]+\frac{(1-p_1)(1-p_2)}{4}\Tr[X^\T\otimes Y^\T\rho_v] \n \\
	 &=&\frac{1-3p_1p_2v}{16}-\frac{v(1-p_1)(1-p_2)(1-2x_0)(1-2y_0)}{16}-\frac{v(x_1y_1+x_2y_2)(1-p_1)(1-p_2)}{4}. \n
\end{eqnarray}
\end{widetext} 
Hence this MDI-EW can identify entangled states if
\begin{eqnarray}
v&>&A^{-1} \text{ [for $A \geq 0$] or } \n \\
v&<&A^{-1} \text{ [for $A < 0$]}, \label{eq10} 
\end{eqnarray}
where $A=[3p_1p_2+(1-p_1)(1-p_2)\{(1-2x_0)(1-2y_0)+4(x_1y_1+x_2y_2)\}]$. Since $0\leq v \leq 1$, we can ignore the second inequality, i.e., the cases of negative $A$. 
The values of $A$, minimized and maximized over all the states $X$ and $Y$, are given by 
$A_{min}=3p_1p_2-(1-p_1)(1-p_2)$ and $A_{max}=3p_1p_2+(1-p_1)(1-p_2)$, respectively. 
In Fig. \ref{fig1}, we plot the lower bounds on $\nu$ corresponding to these two cases in the left and right panels. The left panel corresponds to the case when $A=A_{min}$ and the right panel corresponds to the case of $A=A_{max}$.

\subsection{Bit-flip, bit-phase flip, and phase flip}
The bit-flip, bit-phase flip, and phase flip noises can be represented by using the Pauli matrices (i.e., $\sigma_1$, $\sigma_2$, or $\sigma_3$). For example, if $|0\ket$ and $|1\ket$ are eigenvectors of $\sigma_3$,  $\sigma_1|0\ket=|1\ket$ and $\sigma_1|1\ket=|0\ket$, giving the action of 
the bit flip noise. 

\begin{figure*}[t]
	\includegraphics[scale=1.2]{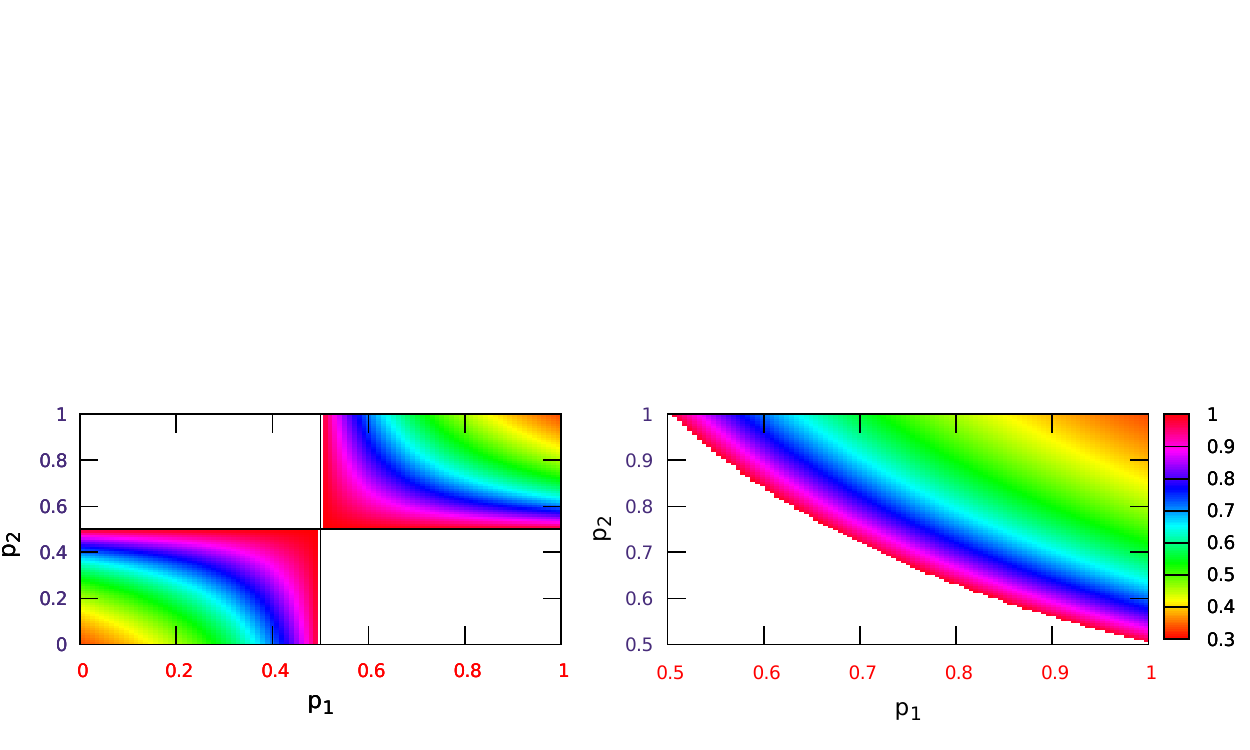}
	\caption{Condition for identification of entangled
Werner state using noisy MDI-EW for bit-flip, bit-phase flip, and phase flip noise models. Even if the input states of the MDI-EW are
admixed with bit flip or bit-phase flip or phase flip noise, with noise parameters $p_1$ and $p_2$, the MDI-EW can spot its entanglement 
if $v$ is above the value indicated by the color. The left  and the right panels correspond to the cases when the both the quantum inputs are affected by same noise (e.g., bit-flip noise on both the inputs) and different noise (e.g., a bit-flip noise on one quantum input and a phase-flip noise on another), respectively. For the case of same noise on both the inputs, noise parameters $p_1$ and $p_2$, together, can take values either greater than \(1/2\) or less than \(1/2\), in order to detect any entangled Werner state, which is illustrated in the left panel by the separation of two colored regions. 
} 
	\label{fig2}
\end{figure*}

Let the noisy input states be such that they are probabilistically acted on by 
a Pauli matrix. The noisy input states under the action of Pauli noise reads as
\begin{equation}
(\ts)_i=p_1\ts+(1-p_1)\sigma_i\ts\sigma_i\text{, }(\ot)_j=p_2\ot+(1-p_2)\sigma_j\ot\sigma_j, \n
\end{equation} 
where $i,j=1,2,3$.  The noisy MDI-EW for Werner states, therefore, becomes
\begin{eqnarray}
	I_{i,j}(P)&=&p_1p_2 I(P)\n \\ &+&\frac{p_1(1-p_2)}{4}\Tr\left[W(I_2\otimes \sigma_j^\T)\rho_v(I_2\otimes\sigma_j^\T)\right]\n \\
	&+&\frac{p_2(1-p_1)}{4}\Tr\left[W\sigma_i^\T\otimes I_2 \rho_v \sigma_i^\T \otimes I_2\right]\n \\
	&+&\frac{(1-p_1)(1-p_2)}{4}\Tr\left[W\sigma_i^\T\otimes\sigma_j^\T\rho_v\sigma_i^\T\otimes\sigma_j^\T\right]. \n
\end{eqnarray} 
Some algebra gives us that
$I_{i,j}(P)=f_1(p_1,p_2)$ for all $i=j$ and $I_{i,j}(P)=f_2(p_1,p_2)$ for all $i\neq j$, where
\begin{eqnarray}
	&&f_1(p_1,p_2)=\frac{1+(4p_1+4p_2-8p_1p_2-3)v}{16} \text{ and} \n \\
	&&f_2(p_1,p_2)=\frac{1+v(1-4p_1p_2)}{16}. \n
\end{eqnarray}
 Therefore if both the input states are affected by the same noise, i.e. $i=j$, then the condition for entanglement detection of an entangled Werner state is
 \begin{eqnarray}
 	v>\frac{1}{8p_1p_2-4p_1-4p_2+3}  \text{ [for positive denominator]},\n \\
 	v<\frac{1}{8p_1p_2-4p_1-4p_2+3}\text{ [for negative denominator]}.\n \\  \label{eq11} 
 \end{eqnarray}
On the other hand, if $i\neq j$, i.e. the two noises in the two input states are different from each other, entanglement in the Werner state is revealed when
\begin{eqnarray}
	v&>&\frac{1}{4p_1p_2-1}\text{ for }p_1p_2>\frac{1}{4}, \n \\
	v&<&\frac{1}{4p_1p_2-1}\text{ for }p_1p_2<\frac{1}{4}. \label{eq12}
\end{eqnarray}
Again, considering the same logic as mentioned for \eqref{eq10}, we can ignore the second inequalities of \eqref{eq11} and \eqref{eq12}.
In Fig. \ref{fig2} we plot the two lower bounds on $\nu$ given in \eqref{eq11}
 and \eqref{eq12}. The left and right panels, respectively, show the values of $\nu$ above which the entangled state can be detected when the two inputs of the MDI-EW  suffer same and different noises. Note that from \eqref{eq11}, we can see that $p_1$ and $p_2$ have to be either less than \(1/2\) or greater than \(1/2\) together, in order to detect any entangled Werner state, which explains the two straight line boundaries dividing the colored regions in the left panel of Fig. \ref{fig2}. 
 
 \subsection{Amplitude-damping noise}
 
\begin{figure}
	\includegraphics[scale=.6]{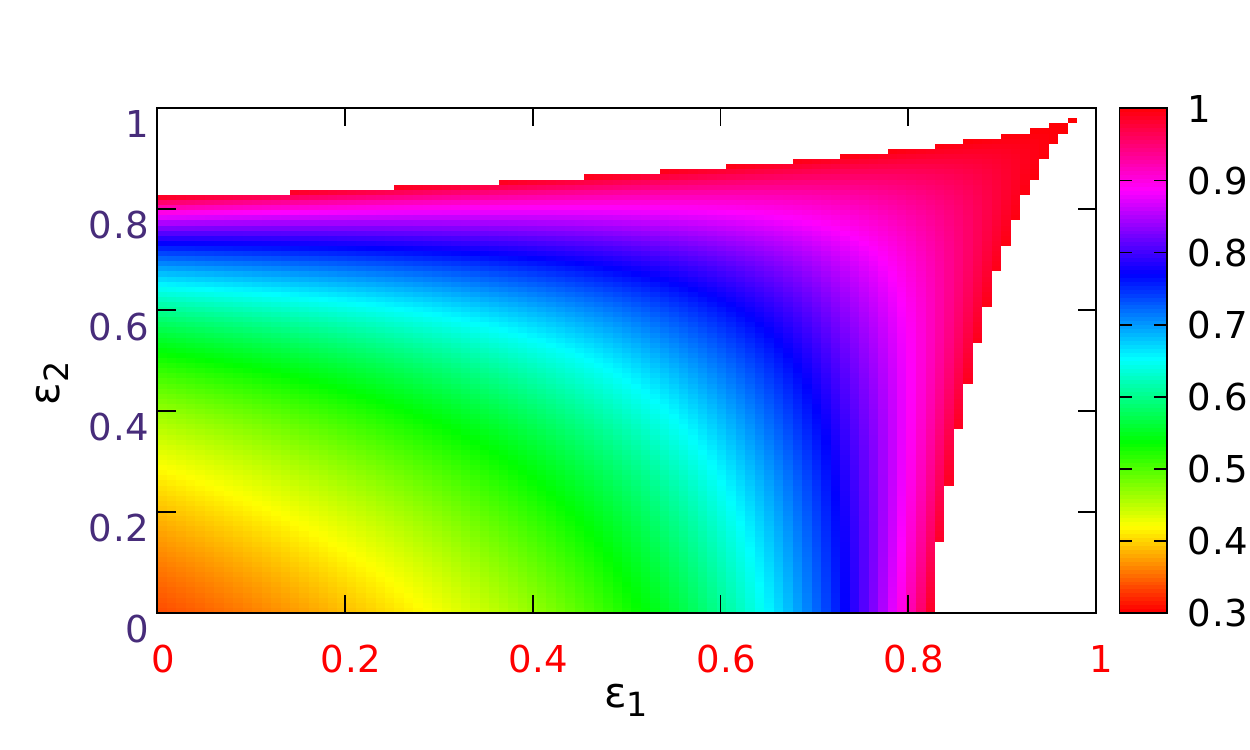}
	\caption{Successful detection of entangled Werner
state using noisy MDI-EW for the amplitude-damping noise in the quantum inputs.  The colored region represents the bounds on
entangled states that can be detected using MDI-EW in presence of
amplitude damping noise with parameters $\epsilon_1$ and $\epsilon_2$. All
quantities are dimensionless.}
	\label{fig3}
\end{figure}

In this subsection, we analyze the effect of noise when both the inputs are affected by local amplitude-damping channels. The input states transform under the channel as
\begin{equation}
	(\ts)_{ad}=\sum_{\mu}M_\mu \ts M_\mu^\dagger\text{, }(\ot)_{ad}=\sum_{\mu}M_\mu \ot M_\mu^\dagger. \n
\end{equation} 
Here, $M_\mu$'s are the amplitude-damping Kraus operators with the property $\sum_{\mu}M_\mu^\dagger M_\mu=I$, and the subscript, $ad$, represents ``amplitude damping''. Due to the presence of such noise, the MDI-EW operator takes the form,

\begin{eqnarray}
	I_{ad}(P)&=&\sum_{s,t,\mu,\nu}\beta_{s,t}\Tr[(M_\mu^\T \ts^\T M_\mu^*)\otimes (M_\nu^\T\ot^\T M_\nu^*)\rho_v]/4\n \\
	&=&\sum_{\mu,\nu}\Tr[W M_\mu^*\otimes M_\nu^*\rho_vM_\mu^\T \otimes M_\nu^\T]/4. \label{eq8}
\end{eqnarray}
The Kraus operators corresponding to the amplitude-damping channel are given by
\begin{eqnarray}
&&	M^a_0=\left(\begin{matrix}
	1&&0\\0&&\sqrt{1-\epsilon}
	\end{matrix}\right)\text{, }
	M^a_1=\left(\begin{matrix}
	0&&\sqrt{\epsilon}\\0&&0
	\end{matrix}\right),\n  
\end{eqnarray}
in the computational basis.
The variable $\epsilon$ is the noise parameter of the  channel. Putting these matrix forms of the Kraus operators in Eq. \eqref{eq8}, and focusing on Werner states, we obtain the following conditions for detecting its entanglement using $I_{ad}(P)$:
\begin{equation}
\nonumber
v>\left[1-\epsilon_1-\epsilon_2+2\epsilon_1\epsilon_2 +2\sqrt{(1-\epsilon_1)(1-\epsilon_2)}\right]^{-1}.  
\end{equation} 
Here 1 or 2 written in the suffixes of the noise parameters indicate noise corresponding to Alice's or Bob's input. In Fig. \ref{fig3}, we plot this bound on $\nu$.

\subsection{Noisy channel with memory}

The channels considered until now are all memoryless, while realistic channels are often not so. 
Let us now consider a scenario where the noises in the two input states are not independent, but correlated by some memory effect \cite{vello'11}. Here, the term ``memory'' is used in the sense that within a single run of the experiment, noise to Bob's input has memory of that to Alice's one, of the same run. This is an example of a global noise. 
Here we focus on Pauli correlated noise \cite{vello'11}. The noisy joint input state under the action of such noise is given by 
\begin{eqnarray}
	&&(\ts\otimes\ot)_{cn}=\sum_{i,j}A_{ij}(\ts\otimes\ot) A_{ij}^\dagger, \n \\
	\text{where } &&A_{ij}=\sqrt{(1-m)p_ip_j+mp_j\delta_{ij}}\sigma_i\otimes\sigma_j\text{ and }i,j=1,2,3.\n
\end{eqnarray}
Here, $0\leq m\leq 1$. The corresponding noisy MDI-EW can be represented as
\begin{equation}
	I_{cn}(P)=\sum_{i,j}\Tr(A_{ij}^\T WA_{ij}^*\rho_v). \n
\end{equation}
Some calculations show 
that the noisy MDI-EW can successfully detect an entangled Werner state if 
\begin{equation}
	v>\frac{1}{3+8(m-1)\sum_{i,j}p_ip_j}. \n
\end{equation} 
Note that 
for $m=1$, 
the condition for entanglement detection of Werner state in the memoryless case is reproduced.

\section{Conclusion}
\label{sec6}
In the laboratory, one of the most appropriate methods to identify entanglement of a shared state is to employ entanglement witnesses (EWs), since there exists an associated locally measurable witness operator for every entangled state, guaranteed by the Hahn-Banach theorem.
In general, standard EWs rely on trusted measurements for the purpose of faithful detection. On the contrary, measurement-device-independent entanglement witnesses (MDI-EWs) never indicate a separable state as entangled even for wrong measurements, and are also robust against a particular kind of lossy detectors. 
Previous works in the literature on MDI-EWs assumed that the ``quantum input'' states, required in MDI-EWs detection process, are perfect. Here, the phrase ``quantum input'' is used for input quantum states that are needed by the observers to carry out the relevant MDI-EW protocol, and these quantum states are different from the shared quantum state whose entanglement is to be detected. Here we relaxed the condition of preparation of perfect quantum inputs and considered noisy inputs within a general framework of noisy MDI-EWs. Noisy inputs may occur due to some imperfections in the preparation devices or errors in the transmission lines. We classified noise processes into two broad classes, uniform and non-uniform, according to their action on the inputs. We found that in the presence of uniform noise over the input states, if the adjoint of the corresponding noise maps separable operators to separable ones, fake entanglement detection never occurs. We also considered examples of non-uniform noise and global entangling noise, where a separable state appear as an entangled one in the identification process, if the measurement device is not trusted. We explored MDI-EWs for the Werner class of states with various noise models on the input states and determined conditions for faithful detection of their entanglement. Finally, we examined an example of correlated noise due to a memory effect, and derived the condition for revealing entanglement of Werner states in this scenario.

\section*{Acknowledgments}
We acknowledge support from the Department of Science and Technology,
Government of India through the QuEST grant (grant numbers
DST/ICPS/QUST/Theme-1/2019/23 and DST/ICPS/QUST/Theme-3/2019/120).

\end{document}